\documentclass[iop]{emulateapj}
\usepackage{natbib,hyperref,verbatim,CJK}
\begin{document}

\title{Be Stars in the Open Cluster NGC\,6830}

\begin{CJK*}{UTF8}{bkai}

\author{
Po-Chieh~Yu (俞伯傑)\altaffilmark{1,2},
Chien-Cheng~Lin (林建爭)\altaffilmark{1,3},
Hsing-Wen~Lin (林省文)\altaffilmark{1},
Chien-De~Lee (李建德)\altaffilmark{1},
Nick~Konidaris\altaffilmark{4},
Chow-Choong~Ngeow (饒兆聰)\altaffilmark{1}, 
Wing-Huen~Ip (葉永烜)\altaffilmark{1,5},
Wen-Ping~Chen (陳文屏)\altaffilmark{1,6},
Hui-Chen~Chen (陳慧真)\altaffilmark{7},
Matthew~A.~Malkan\altaffilmark{2},
Chan-Kao~Chang (章展誥)\altaffilmark{1},
Russ~Laher (良主嶺亞)\altaffilmark{8},
Li-Ching~Huang (黃立晴)\altaffilmark{1},
Yu-Chi~Cheng (鄭宇棋)\altaffilmark{1},
Rick~Edelson\altaffilmark{9},
Andreas~Ritter\altaffilmark{1,10},
Robert~Quimby\altaffilmark{11},
Sagi~Ben-Ami\altaffilmark{12},
Eran.~O.~Ofek\altaffilmark{12},
Jason Surace\altaffilmark{8}, and
Shrinivas~R.~Kulkarni\altaffilmark{13}
}

\altaffiltext{1}{Graduate Institute of Astronomy, National Central University, 300 Jhongda Road, Jhongli 32001, Taiwan}
\altaffiltext{2}{Department of Physics and Astronomy, University of California, Los Angeles, CA 90024, USA}
\altaffiltext{3}{Key Laboratory for Research in Galaxies and Cosmology, Shanghai Astronomical Observatory, Chinese Academy of Sciences, 80 Nandan Road Shanghai 200030, China}
\altaffiltext{4}{Cahill Center for Astronomy and Astrophysics, California Institute of Technology, 1200 E California Blvd., Pasadena, CA 91125, USA}
\altaffiltext{5}{Institute of Space Science, National Central University, 300 Jhongda Road, Jhongli 32001, Taiwan}
\altaffiltext{6}{Department of Physics, National Central University, 300 Jhongda Road, Jhongli 32001, Taiwan}
\altaffiltext{7}{Department of Natural Sciences and Sustainable Development, Ministry of Science and Technology, 106, Sec. 2, Heping E. Rd., Taipei 10622, Taiwan}
\altaffiltext{8}{Spitzer Science Center, California Institute of Technology, M/S 314-6, Pasadena, CA 91125, USA} 
\altaffiltext{9}{Department of Astronomy, University of Maryland, College Park, MD 20742, USA}
\altaffiltext{10}{Department of Astrophysical Sciences, Princeton University, Princeton, NJ 08544, USA}
\altaffiltext{11}{Kavli-Institute for the Physics and Mathematics of the Universe, University of Tokyo, Kashiwanoha 5-1-5, Kashiwa-shi, Chiba, Japan}
\altaffiltext{12}{Department of Particle Physics and Astrophysics, The Weizmann Institute of Science, Rehovot 76100, Israel}
\altaffiltext{13}{Division of Physics, Mathematics and Astronomy, California Institute of Technology, Pasadena, CA 91125, USA} 

\begin{abstract}
We report the discovery of 2 new Be stars, and re-identify one known Be star in the open cluster NGC\,6830.
Eleven H$\alpha$ emitters were discovered using the H$\alpha$ imaging photometry of the Palomar Transient Factory Survey.
Stellar membership of the candidates was verified with photometric and kinematic information using 2MASS data and proper motions.
The spectroscopic confirmation was carried out by using the Shane 3-m telescope at Lick observatory.
Based on their spectral types, three H$\alpha$ emitters were confirmed as Be stars with H$\alpha$ equivalent widths $>$ $-$10\AA.
Two objects were also observed by the new spectrograph SED-Machine on the Palomar 60 inch Telescope.
The SED-Machine results show strong H$\alpha$ emission lines, which are consistent with the results of the Lick observations.
The high efficiency of the SED-Machine can provide rapid observations for Be stars in a comprehensive survey in the future.
\end{abstract}


\section{Introduction}
Be stars are non-supergiant B-type stars with Balmer emission lines, especially the H$\alpha$ lines.
Be stars are also characterized by their fast rotation, color excess at infrared wavelength, and continuum/emission-line variability \citep{tow04,Koubsky1997,Hubert1998}.
Some Be stars show the variability with periods that might be due to pulsation or rotation \citep{Porter2003,Rivinius2003}.
Previous studies have found the enhancement of Be phenomenon in young clusters \citep{wis06,mat08}, while \citet{McSwain05} showed an overall increase
in Be frequency with age until 100 Myr. \citet{McSwain05} also indicated that Be star formation is not strongly related to the cluster density.
Furthermore, the Be star fraction in open clusters seems to be increasing with lower metallicity;
between 17.5\% and 40\% of B stars were found to be Be stars in the Large Magellanic Cloud  (LMC) and Small Magellanic Cloud (SMC)  \citep[see detailed review in][]{Rivinius2013}.

It is well accepted that the H$\alpha$ emissions of Be stars originate from the circumstellar material, which is suggested to be a flattened disk \citep{Carciofi2006,Quirrenbach1994,Quirrenbach1997,Wood1997}.
The disks are found to be in Keplerian rotation and stable \citep{Kraus2012,Meilland2012}. 
The material might be moved from the stellar surface to form the circumstellar disk (e.g., the decretion disk) through the combination of fast rotation and other mechanisms, such as nonradial pulsations or magnetic fields \citep{Lee1991,Porter2003,tow04,Carciofi2009}.
There are several possible origins of the fast rotating star: (1) binary interaction \citep{Pols1991}. During the interaction process, mass and angular momentum can
be transferred from the primary star to the mass gainer star, and the latter might be spun-up to a fast rotating star; (2) they might be born as fast rotators; and (3) they have been spun-up
during the Main-Sequence (MS) evolution.

\begin{figure*}
\epsscale{1}
\plotone{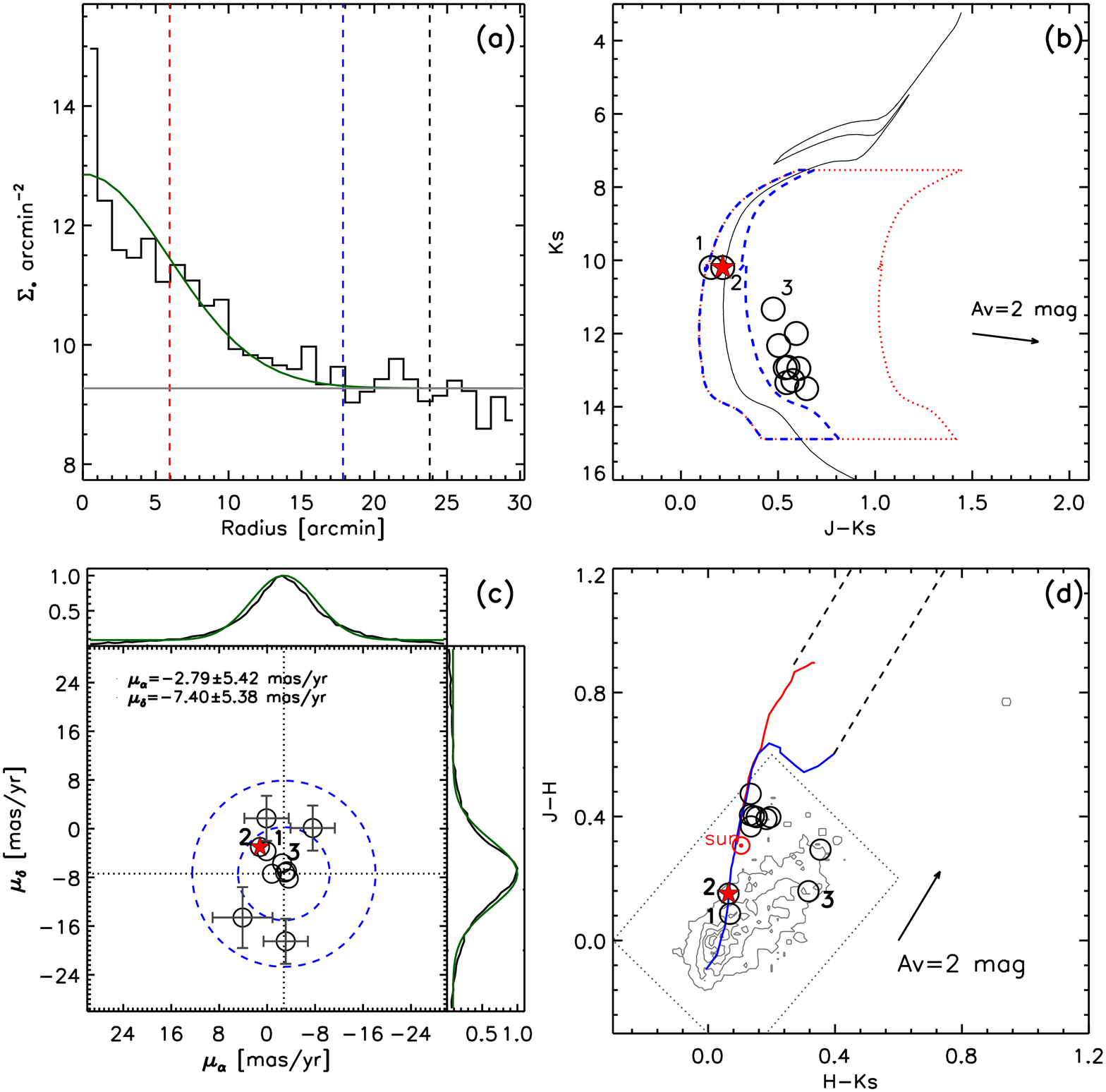}
\caption{Membership identification. In Figure~(b) -- (d), open circles represent our Be star candidates, while the red pentagram is the known Be star VES\,72. 
Numbers represent spectroscopically confirmed Be stars NGC6830-1, NGC6830-2, and NGC6830-3 in this study.
(a) Radial density profile of NGC\,6830. The gray horizontal line indicates the background density. The green curve show the best Gaussian fitting results while red and blue vertical lines represent the width of 1$\sigma$ and 3$\sigma$. The
black vertical line shows the width of the Gaussian curve of 4$\sigma$. (b) 2MASS color-magnitude diagram. The black solid line indicates the isochrone \citep{gir02} of 125 Mys located at 1.8 kpc. Be candidates are considered to have photometric membership within the red-dashed region, while non-emission-line stars are considered to have photometric membership within the blue-dashed region. (c) Stellar PMs distribution of Be candidates of NGC\,6830. The PM of NGC\,6830 is ($\mu_{\alpha}$,$\mu_{\delta}$) = ($-$2.79,$-$7.40) mas year$^{-1}$. The inner and outer blue-dashed circles represent 1$\sigma_{\mu}$ and 2$\sigma_{\mu}$ regions. We adopted 2$\sigma$ region to define the kinematic membership. The bars are errors of the PMs. The upper and right panels show the $\mu_{\alpha}$ and $\mu_{\delta}$ of stars within 48$\arcmin$ $\times$ 48$\arcmin$; green lines show the Gaussian fitting to PMs. (d) 2MASS color-color magnitude diagram. The gray contours show known Be stars distribution, and we define dashed box to select possible Be candidates.}
\label{all}
\end{figure*}

To further understand the nature of Be stars, several surveys of Be stars have been carried out in the past \citep*{Mclau1932, Jaschek1964, Drew2005, Drew2015, Raddi2015,Lin2015}.
Because stars in clusters share age and metallicity, Be stars in a sample of clusters with different ages play an essential role in the study of their evolution.
Previous studies have searched for emission-line objects in a number of star clusters in the SMC and MW.
\citep*{fab00,kel01,wis06,Martayan2010,mat08,mathew11}.
Particularly, \citet{McSwain05} provided the most extensive survey of Be stars in the MW;
they investigated 55 clusters and concluded that $73\%$ of the rapid rotators might be spun-up by the effect of mass transfer between binaries. 
Moreover, \citet{McSwain09} also showed that the distributions of projected rotational velocity of Be stars and normal B-type stars are significantly different, indicating that
they might be different stellar populations, and Be stars cannot be drawn from a sample of normal B stars.
Theoretical simulations also imply that most Be stars could be formed during the phase of binary interactions \citep{deMink2013,Shao2014}.
These studies further supported a spin-up scheme for the evolutionary sequence of Be stars.

Nevertheless, the sample of Be stars in open clusters is far from complete due to several difficulties:
(1) a comprehensive spectroscopic survey is time consuming, (2) spectroscopic surveys are often limited to bright stars.
Therefore, with the wide field of view (7.3 square degree) of the 48 inch Samuel Oschin Telescope, the Palomar Transient Factory \citep[PTF;][]{Law2009} project provides
an efficient way to search for emission-line candidates in open clusters using H$\alpha$ imaging photometry \citep{Yu2015}.
Furthermore, a new instrument called the SED-Machine \citep*{Sagi2012, Ngeow2013, Ritter2014} provides a possible solution for the fast and efficient follow-up spectroscopic 
observations for the emission-line candidates found in PTF, because the highly efficient SED-Machine is designed to classify transient events rapidly with
a low resolution ($\lambda$/$\delta\lambda$$\sim$100) integrated field unit (IFU) spectrograph that covers a large wavelength range (370 to 920~nm).
Thus, a comprehensive survey of Be stars in star clusters can be accomplished with a combination of the PTF H$\alpha$ imaging photometry and the SED-Machine.

Here we present the results of searching for Be stars in an open cluster NGC\,6830 ($\alpha_{2000}=19^{\rm h}50^{\rm m}59^{\rm s}, \delta_{2000}=23\degr06\arcmin00\arcsec$).
NGC\,6830, located at a distance of 1.8~kpc \citep{kha13} in the constellation Vulpecula \citep{Barkhatova1957}, has a fairly loose structure.
In contrast to our previous work \citep[NGC 663 with an age of $\sim31$~Myrs,][]{Yu2015}, the age of NGC\,6830 is about 125~Myrs \citep{kha13} and only one Be star has been reported in this cluster \citep{Hoag1965}
prior to our work. Using the relation of effective temperature and age, \citet{Pena2011} determined that the age of NGC\,6830 is log(age) = 7.69 yr (48.9 Myrs) with the hottest star (17,000 K) in the membership of 19 stars.
It seems that the age given by \citet{kha13} is not consistent with that of \citet{Pena2011}.
However, while the membership in \citet{Pena2011} is only determined from the distance, \citet{kha13} included the proper motions (PMs) to determine the membership and give the log(age)  = 8.105$\pm$0.062 yr (125~Myrs).
Furthermore, the range of the effective temperature for the membership in \citet{Pena2011} is from 12,000 to 17,000 K, indicating the log(age) is between 7.69 and 8.22 yr, which is also consistent with the age given by \citet{kha13}.
Thus, here we adopted the age given by \citet{kha13}.
 
In addition to the age difference, we selected NGC\,6830 as the subject of this work because the known Be star is fairly bright and hence it is suitable to be observed with the SED-Machine during its commissioning runs. 
Therefore, the main goal of this work is to demonstrate the capability of the combined PTF H$\alpha$ imaging photometry, 
the archival 2MASS and proper motions data, and the SED-Machine in the search of Be stars in open clusters. This will lay the foundation of our future work in this area. 
In Section~2, we describe the methodology of identifying emission-line stars and determination of cluster membership using the PTF H$\alpha$ imaging photometry.  
In Section~3 and~4, we describe our spectroscopic observations and present our results based on these observations. 
Section~5 gives a summary and discussion of this study.

\begin{figure*}[!ht]
\centering
\epsscale{0.8}
\includegraphics[angle=0]{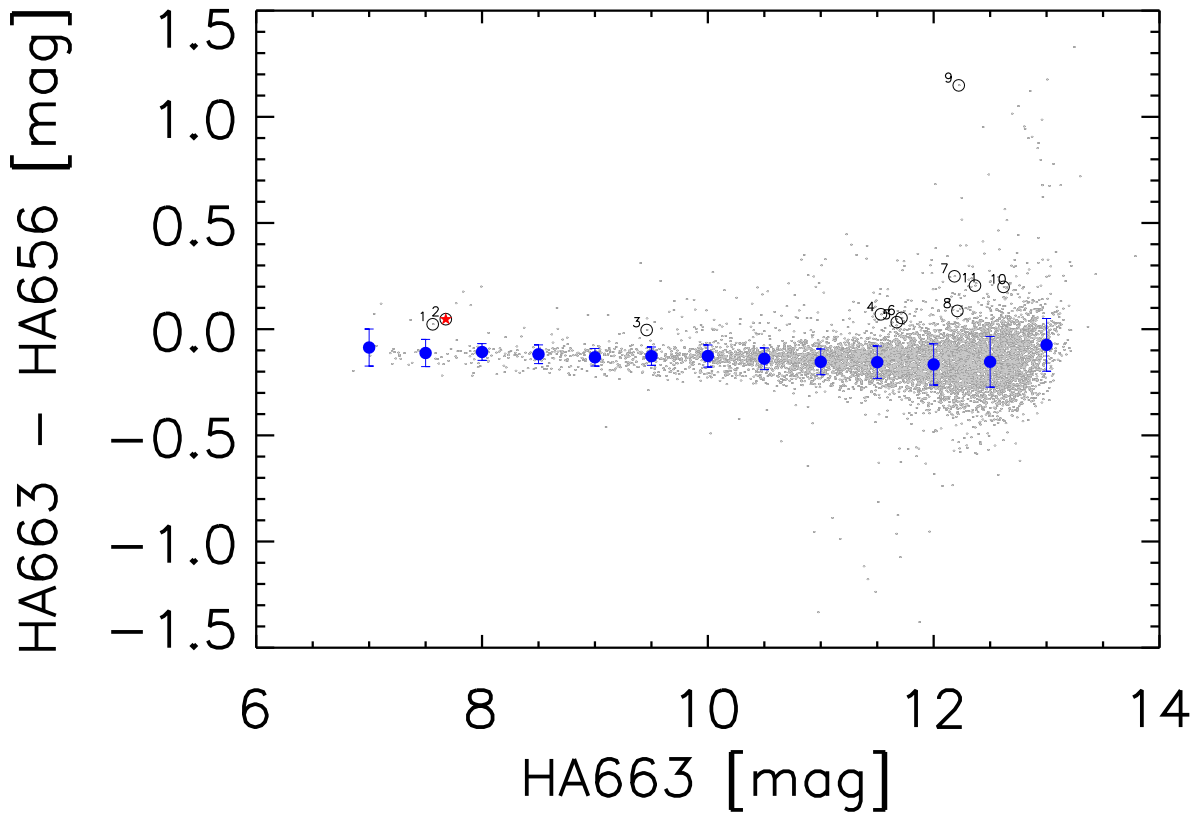}
\caption{H$\alpha$ emitter candidates selection.
Gray dots: stars used for calculating photometric scattering; open circles: emission-line candidates with membership confirmation; numbers represent Be candidates NGC6830-1 -- NGC6830-11; 
red pentagram: known Be stars VES\,72. Blue filled circles represent
mean values of HA663$-$HA656 values within each 0.5 mag bin. The error bars represent the photometric scattering $\sigma_{p}$, which are calculated with the error propagated from photometric and systematic errors.
Stars with HA663$-$HA656 $>$ mean+2$\sigma_{p}$ are selected as emission-line candidates.}
\label{HA656}
\end{figure*}

\begin{deluxetable*}{lrrrrrrrrrr}
\tabletypesize{\small}
\tablecolumns{11} \tablewidth{0.99\textwidth}\tablecaption{Be Candidates}
\tablehead{
\colhead{ID} &
\colhead{R.A.} &
\colhead{Decl.} &
\colhead{$\mu_{\alpha}$} &
\colhead{$\mu_{\delta}$} &
\colhead{epm} &
\colhead{$K_{s}$} &
\colhead{$J{-}H$} &
\colhead{$H{-}K_{s}$} &
\colhead{HA663} &
\colhead{HA663$-$HA656}\\
\colhead{} &
\colhead{(hh:mm:ss)} &
\colhead{(hh:mm:ss)} &
\colhead{(mas~yr$^{-1}$)} &
\colhead{(mas~yr$^{-1}$)} &
\colhead{(mas~yr$^{-1}$)} &
\colhead{(mag)} &
\colhead{(mag)} &
\colhead{(mag)} &
\colhead{(mag)} &
\colhead{(mag)}
}
\startdata
NGC\,6830-1 &19:51:00.47 &+23:06:50.6       &0.10    &$-$3.70    &1.50 &10.350  &0.086  &0.069   &7.563  &0.023\\
NGC\,6830-2 &19:50:58.55 &+23:08:28.4       &1.20    &$-$3.00    &1.50 &10.412  &0.151  &0.064   &7.679  &0.047\\
NGC\,6830-3 &19:51:58.13 &+22:44:59.8       &$-$3.30    &$-$7.00    &3.70 &11.806  &0.159  &0.316   &9.458 &$-$0.004\\
NGC\,6830-4 &19:50:02.31 &+23:00:15.2       &$-$3.60    &$-$8.20    &3.70 &12.831  &0.368  &0.135  &11.533  &0.070\\
NGC\,6830-5 &19:50:05.02 &+23:02:55.2     &$-$2.50    &$-$5.70    &3.70 &12.592  &0.398  &0.196  &11.673  &0.033\\
NGC\,6830-6 &19:50:04.61 &+23:20:27.3     &$-$7.60     &0.10    &3.70 &13.473  &0.405  &0.132  &11.715  &0.053\\
NGC\,6830-7 &19:52:08.58 &+22:59:59.8      &4.10   &$-$14.60    &5.00 &14.151  &0.293  &0.354  &12.184  &0.249\\
NGC\,6830-8 &19:50:02.94 &+23:02:40.2    &$-$3.00    &$-$7.20    &3.70 &13.474  &0.398  &0.153  &12.210  &0.086\\
NGC\,6830-9 &19:52:10.97 &+23:17:06.0     &$-$0.80    &$-$7.40    &3.70 &13.562  &0.473  &0.134  &12.222  &1.148\\
NGC\,6830-10 &19:52:25.39 &+22:49:11.6      &0.10     &1.70    &3.70 &13.886  &0.401  &0.144  &12.619  &0.199\\
NGC\,6830-11 &19:51:07.65 &+23:24:32.6     &$-$3.10   &$-$18.50    &3.70 &13.859  &0.392  &0.184  &12.366  &0.205
\enddata
\tablecomments{Column 1: ID of candidates. Column 2: R.A. in sexagesimal format (J2000). Column 3: decl. in sexagesimal format (J2000). Column 4: proper motion of R.A.
Column 5: proper motion of decl. Column 6: PM uncertainties. 
Column 7: $K_{s}$-band magnitudes adopted from 2MASS. Column 8: $J{-}H$. Column 9: $H{-}K_{s}$.
Column 10: instrumental magnitude of the HA663 filter. Column 11: HA663$-$HA656 magnitude difference.}
\end{deluxetable*}

\begin{deluxetable*}{lrrrrrr}
\tabletypesize{\small}
\tablecolumns{5} \tablewidth{0pt}\tablecaption{Observation Log and Results}
\tablehead{
\colhead{ID} &
\colhead{Date} &
\colhead{Telescope} &
\colhead{Instrument} &
\colhead{Time}&
\colhead{R-band}&
\colhead{Type}
}
\startdata
NGC\,6830-1  &2014/May/1 &Palomar 60 inch &SED Machine &300s &11.19 &Be\\
NGC\,6830-2  &2014/May/1 &Palomar 60 inch &SED Machine &300s &11.51 &Be\\ 
NGC\,6830-1  &2014/Aug/2 &Lulin One-meter &HIYOYU &1200s &11.19 &Be\\
NGC\,6830-1  &2015/June/15 &Lick 3-meter &Kast Dual Spectrograph &600s &11.19 &Be\\
NGC\,6830-2  &2015/June/15 &Lick 3-meter &Kast Dual Spectrograph &600s &11.51 &Be\\
NGC\,6830-3  &2015/June/15 &Lick 3-meter &Kast Dual Spectrograph &1200s &13.24 &Be\\
NGC\,6830-4  &2015/June/15 &Lick 3-meter &Kast Dual Spectrograph &1200s &15.28 &A\\
NGC\,6830-5  &2015/June/15 &Lick 3-meter &Kast Dual Spectrograph &1200s &14.98 &G\\
NGC\,6830-6  &2015/June/16 &Lick 3-meter &Kast Dual Spectrograph &1500s &14.98 &G\\
NGC\,6830-7  &2015/June/16 &Lick 3-meter &Kast Dual Spectrograph &1500s &15.59 &G\\
NGC\,6830-8  &2015/June/15 &Lick 3-meter &Kast Dual Spectrograph &1200s &15.69 &G\\
NGC\,6830-9  &2015/June/16 &Lick 3-meter &Kast Dual Spectrograph &1500s &15.89 &G\\
NGC\,6830-10  &2015/June/16 &Lick 3-meter &Kast Dual Spectrograph &1200s &16.18 &G\\
NGC\,6830-11  &2015/June/16 &Lick 3-meter &Kast Dual Spectrograph &1500s &15.93 &G
\enddata
\tablecomments{Column 1: ID of targets. Column 2: Observations dates. Column 3: Telescope name. Column 4: Instrument name. Column 5: Exposure time in seconds. Column 6: R-band magnitudes adapted from PPMXL catalog.
Column 7: Spectral types.}
\end{deluxetable*}

\section{Identification of Be Candidates and Membership}
The methodology of searching for Be star candidates and identifying membership has been presented in our pilot project. Further details can be found in \citet{Yu2015} and will not be repeated here. 
Briefly, procedures that we followed in this study are summarized below:

(1) Determining the searching region based on the radial density profile. 
The stars were selected from the PPMXL data set with the S/N $\geq$ 10 in 2MASS $J$, $H$, and $K_{s}$ bands.
The half-Gaussian fitting gives a 3$\sigma$ radius of 17.85$\arcmin$  (Figure 1a). 
Because open clusters have irregular shapes, we therefore
adopted a box with the side of 48$\arcmin$ (4$\sigma$) as our searching region for NGC\,6830 to cover possible candidates. 

(2) Applying the H$\alpha$ imaging photometry to identify possible H$\alpha$ emitters. 
The H$\alpha$ emitters should have a significant flux excess in the on-line image than in the off-line image,
where the on-line and off-line images were taken through the HA 656 and HA 663 narrow-band filters.
These images were processed for bias corrections, flat fielding, and astrometric calibration with pipelines developed at the Infrared Processing and Analysis Center \citep[IPAC;][]{Laher2014}.
Following the methods and criteria as given in \citet{Yu2015}, we selected the H$\alpha$ emitter candidates that stand out in the plot of the differential HA663$-$HA656 flux as possible H$\alpha$ emitters (see Figure~2).

(3) Verifying the photometric membership using near-infrared data.
The near-infrared $J$, $H$, and $K_{s}$ bands data are obtained from the 2MASS point source catalog \citep{cut03}.
Possible photometric members of NGC\,6830 are determined by selecting stars within the region that are near the isochrone in the 2MASS $K_{s}$ versus $J-K_{s}$ color-magnitude diagram 
(Figure 1b). 
Considering that Be stars might exhibit a large infrared excess \citep{lee11}, we extended the selection region to $J-K_{s}$ $\sim$ 1.2 mag (Figure~1b) for the H$\alpha$ emitter candidates.

(4) Verifying the kinematic membership using the PMs.
Because about 10\% objects in PPMXL data include spurious entries \citep{ros2010}, \citet{kha12} have averaged their PMs and computed errors. 
Thus, we used PMs provided by \citet{kha12} to calculate the averaged PM ($\mu_{\alpha}$,$\mu_{\delta}$) and standard deviations ($\sigma_{\mu\alpha}$, $\sigma_{\mu\delta}$) of stars within the 47.6$\arcmin$~$\times$~47.6$\arcmin$
region by fitting a Gaussian distribution to the PMs. The adopted $\sigma_{\mu}$ included the error propagated from $\sigma_{\mu\alpha}$ and $\sigma_{\mu\delta}$. 
Stars were then considered as the kinematic membership if their PMs are within the 2$\sigma_{\mu}$ region (Figure~1c).

(5) Selecting possible Be stars from emission-line candidates based on $J-H$ versus $H-K_{s}$ color-color diagram (CMD).
The gray contours as shown in Figure~1d represent the region of 1185 known Be stars \citep{zha05}. 
Under the assumption that most Be stars have similar infrared colors, we identified Be star candidates as those in NGC\,6830 inside the gray-dashed region in the CMD.

Using the above criteria, we identified 11 Be star candidates (Table 1). One candidate was confirmed as the known Be star VES\,72 in the cluster NGC\,6830 \citep[listed as NGC\,6830-2 in Table 1;][]{Hoag1965}.

\section{Spectroscopic Observations and Data Reduction}

\subsection{SED-Machine Observations}
To demonstrate the feasibility and capability of the SED-Machine in the study of Be stars, we obtained the optical spectra of two bright objects, VES\,72 (NGC\,6830-2) and a new Be star candidate NGC\,6830-1 during its commission runs. 
Logs of observations are listed in Table 2.
Due to time constraints during the commission runs, we only observed the two brightest candidates but not all of our 11 candidates. 
The SED-Machine, mounted on the Palomar 1.5-m telescope, observations were carried out on 2014 May 1 under a seeing of approximately 2$\arcsec$. 

The SED-Machine IFU spectra were reduced independently with the pipelines developed by the NCU and the Caltech PTF team. The IFU data reduction procedures consist of the following:
(1) Overscan and bias subtraction; (2) Automatically spectrum identification using a Xe arc-lamp spectra image; 
(3) Calculating wavelength calibration solutions of every IFU spectra using HgNeXe lamp sources;
(4) Simple-sum extraction of all spectra; 
(5) Reconstructing the image of the sky as seen before prism from surrounding sky identification;
(6) Object spectrum extraction with sky spectra subtraction, wavelength calibration and flux calibration.

\subsection{LOT Observations}

To confirm the H$\alpha$ detection from SED-Machine observations, we re-observed NGC\,6830-1 using the Hiyoyu spectrograph on the Lulin One-meter Telescope (LOT). 
The LOT observations were conducted on 2014 Aug 2 under a seeing of approximately 1.5$\arcsec$. 
Using a grating of 300 mm$^{-1}$ and a slit width of 2\arcsec, we covered a wavelength range of 3800$-$7600\AA~. 
The Hiyoyu spectrum was reduced with the IRAF\footnote[1]{Image Reduction and Analysis Facility. IRAF is written and supported by the National Optical Astronomy Observatories (NOAO) in Tucson, Arizona.} 
package, following the standard data reduction procedure,
i.e., dark, bias-subtraced, flat-fielding-corrected, wavelength calibration, and flux calibration. The wavelength calibration 
was performed using a HeNeAr lamp. The flux calibration might be not very accurate due to the unstable weather during the observation.

\subsection{Lick Observations}
We used the Kast dual spectrograph on the 3-m Shane telescope at Lick observatory to complete the confirmation of spectral typing of the 11 Be star candidates.
The observing runs were conducted on 2015 June 15 -- 16 under a seeing of 1.5\arcsec~-- 2\arcsec~using a 2\arcsec~slit width.
Using 600/7500 grating on the red side and 600/4310 grism on the blue side, we can cover a wavelength range of 3500$-$7800\AA~.
The dispersion on the red side and blue side is 2.32 \AA/pixel and 1.85 \AA/pixel, respectively. 
We reduced the spectra using the standard tools in IRAF, including corrections for overscan bias and flat-fielding using the dome flat images. 
We used NeHg-Cd arc lamps for wavelength calibrations, and standard stars Feige\,92 and BD+28~4211 for flux calibration.

\section{Results}

\subsection{SED-Machine and LOT Spectra}
In Figure~3 we showed the SED-Machine spectra of the bright Be stars NGC\,6830-1 and VES\,72 (NGC\,6830-2).
The spectrum of VES\,72 shows prominent H$\alpha$ emission line, and  clear hydrogen and \ion{He}{1}$\lambda$4471 + \ion{Mg}{2}$\lambda$4481 absorption lines.
We further observed the new Be candidate NGC\,6830-1 using the SED-Machine, and the Hiyoyu spectrograph mounted on the LOT.
Both SED-Machine and LOT spectra present a clear detection of the H$\alpha$ emission line, and hydrogen and \ion{He}{1}$\lambda$4471 + \ion{Mg}{2}$\lambda$4481 absorption lines,
indicating a Be type. With a spectral resolution $\lambda$/$\delta\lambda$$\sim$333 of the Hiyoyu spectragraph, we estimated an equivalent width of H$\alpha$ emission-line, EW[H$\alpha$] $= -5$\AA~for NGC\,6830-1.
Although the spectral resolution is too poor to classify the spectral subtype, the similar brightness of NGC\,6830-1 to VES\,72
suggests similar subtypes (e.g. B6 type). 

\begin{figure}[!ht]
\epsscale{1.0}
\plotone{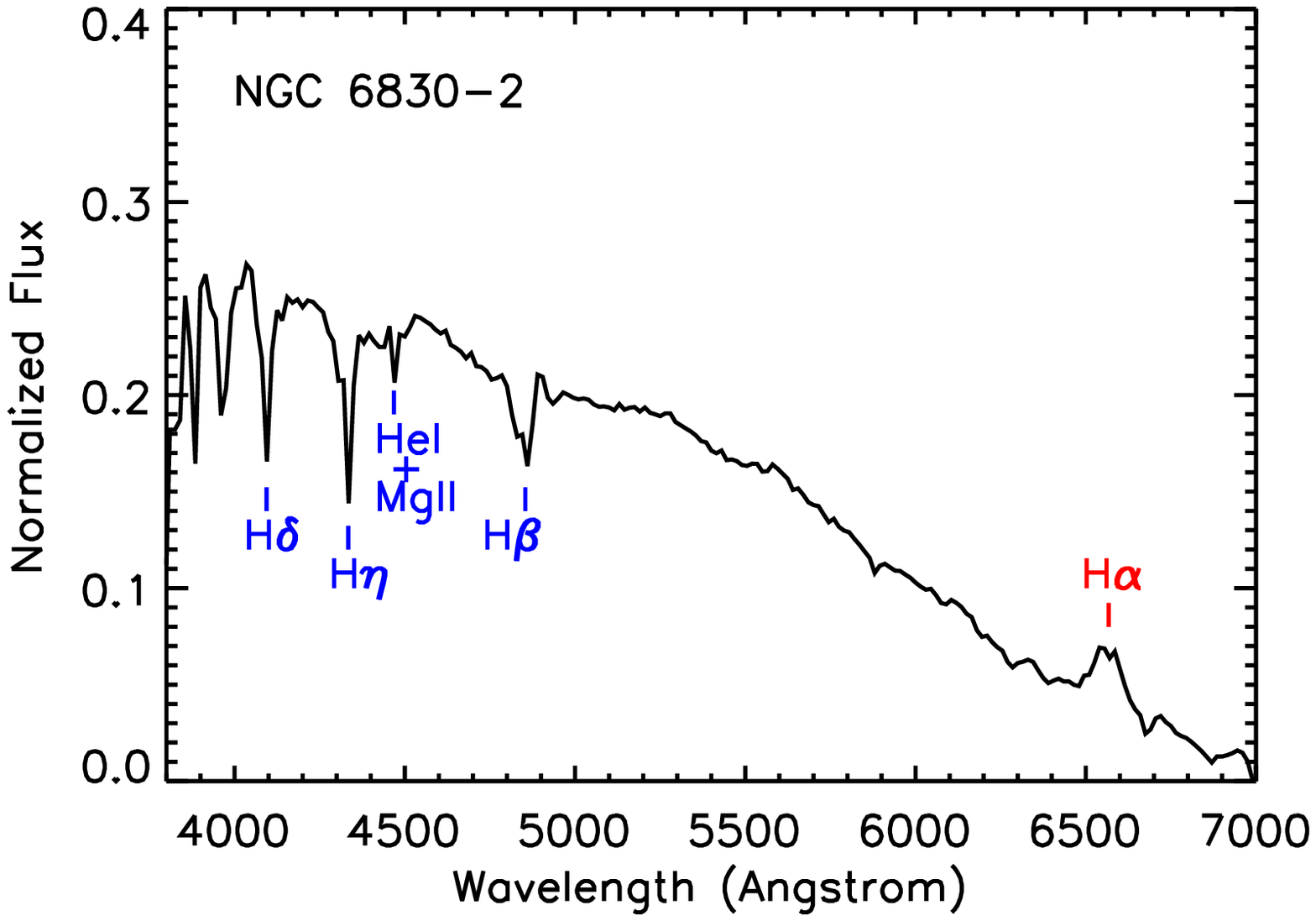}
\plotone{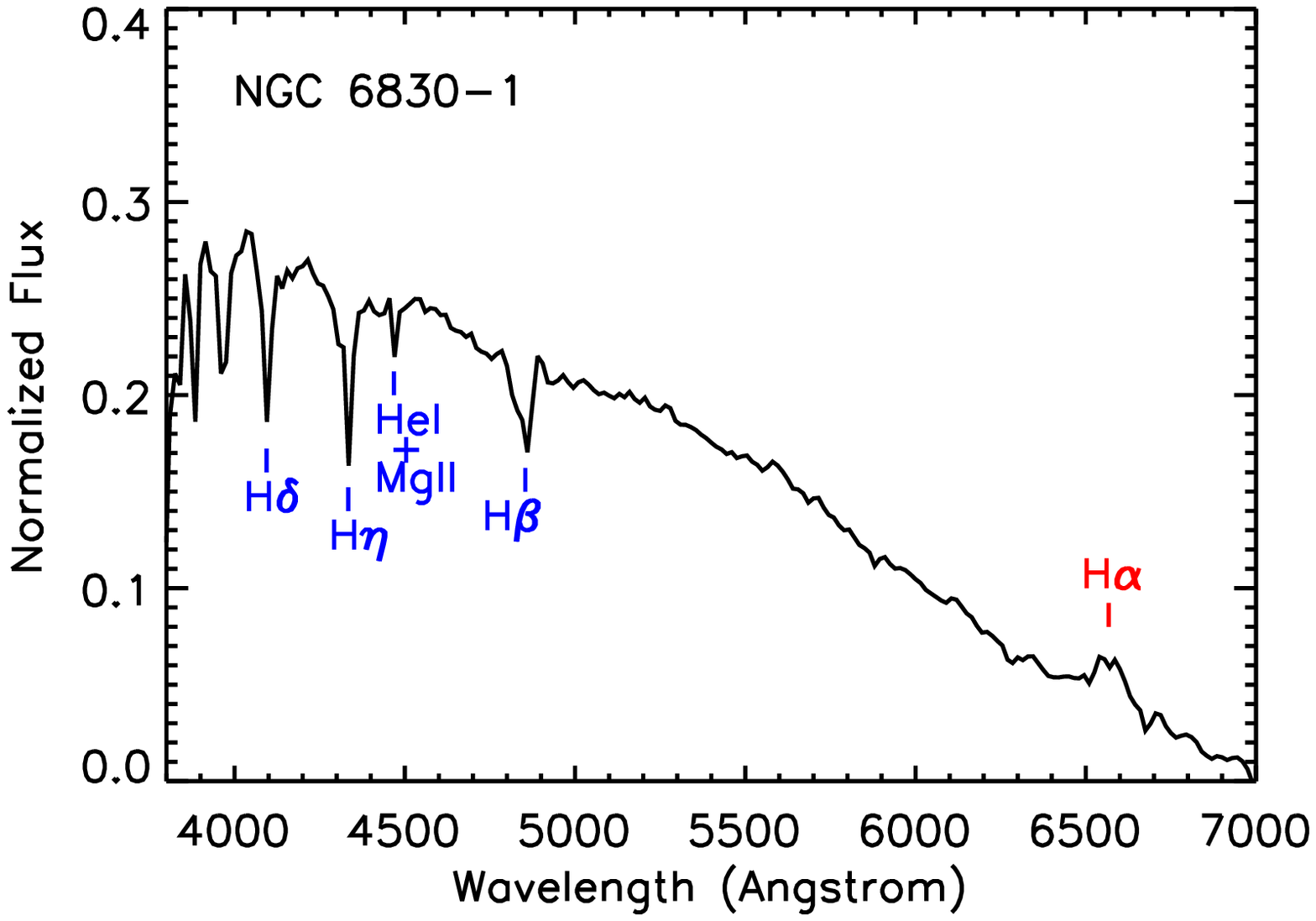}
\plotone{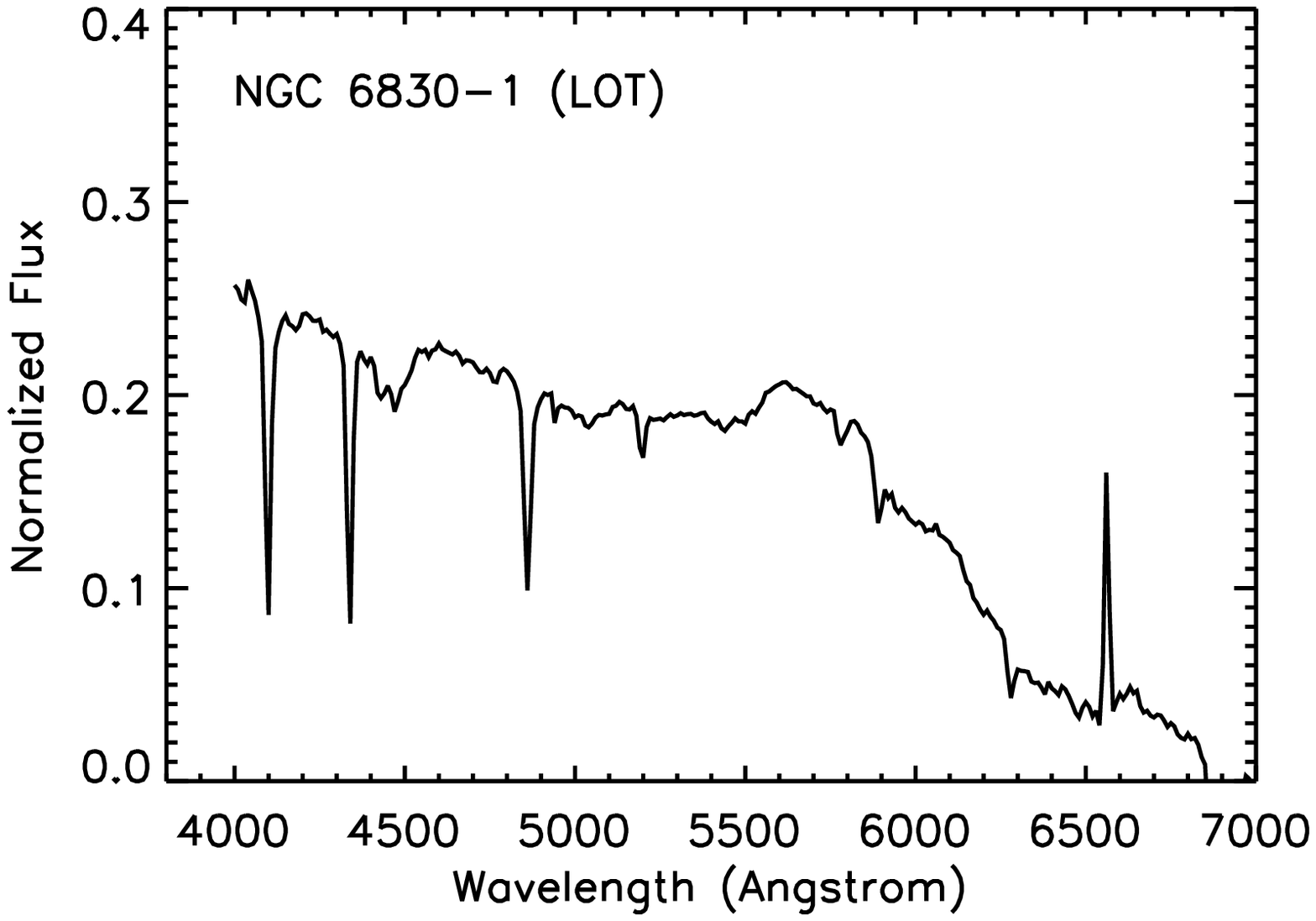}
\caption{SED-Machine results. The unit of the y-axis is normalized fluxes at 7000\AA in logarithm. Upper panel: SED-Machine spectrum of the known Be star VES\,72. We detect the clear H$\alpha$ emission-line while other Balmer series show absorption features. Middle panel: We also detected a clear H$\alpha$ emission-line for the Be star candidate NGC6830-1 with the SED-Machine. Lower panel: The H$\alpha$ emission-line also appears in the LOT spectrum for the same candidate.}
\label{SED-new}
\end{figure}

\subsection{Lick Spectra}
We present the spectra of Be star candidates obtained by Lick 3-m Shane telescope in Figure~4.
Only three Be star candidates (NGC6830-1, NGC6830-2, and NGC6830-3) show H$\alpha$ emission and B-type spectra.
All candidates are discussed in the following subsections.

\begin{figure*}
\centering
\epsscale{1}
\includegraphics[angle=0]{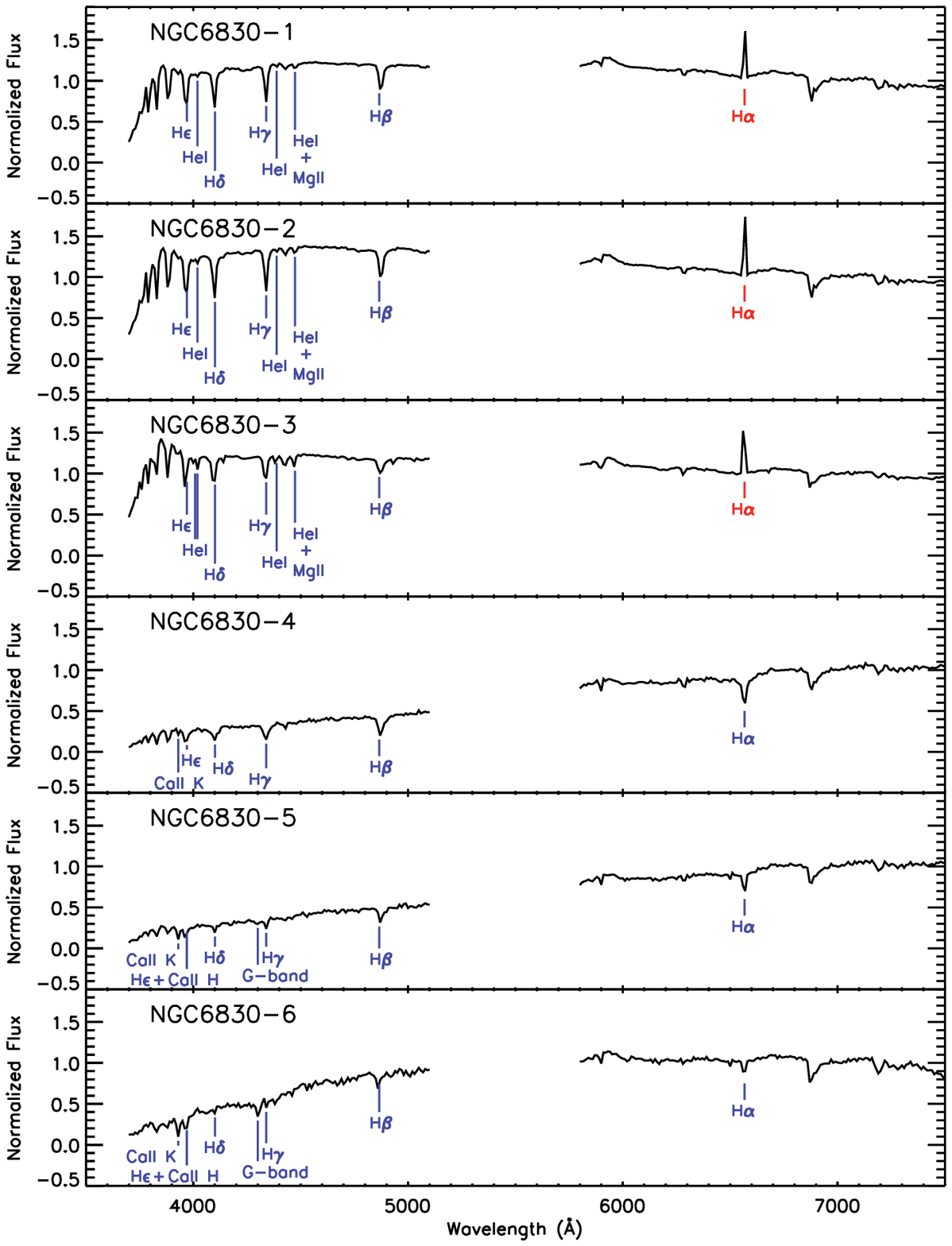}
\caption{Lick spectra of Be star candidates. Fluxes are normalized at 7000\AA.}
\label{Lick1}
\end{figure*}

\renewcommand{\thefigure}{\arabic{figure} (Cont.)}
\addtocounter{figure}{-1}

\begin{figure*}
\centering
\epsscale{1}
\includegraphics[angle=0]{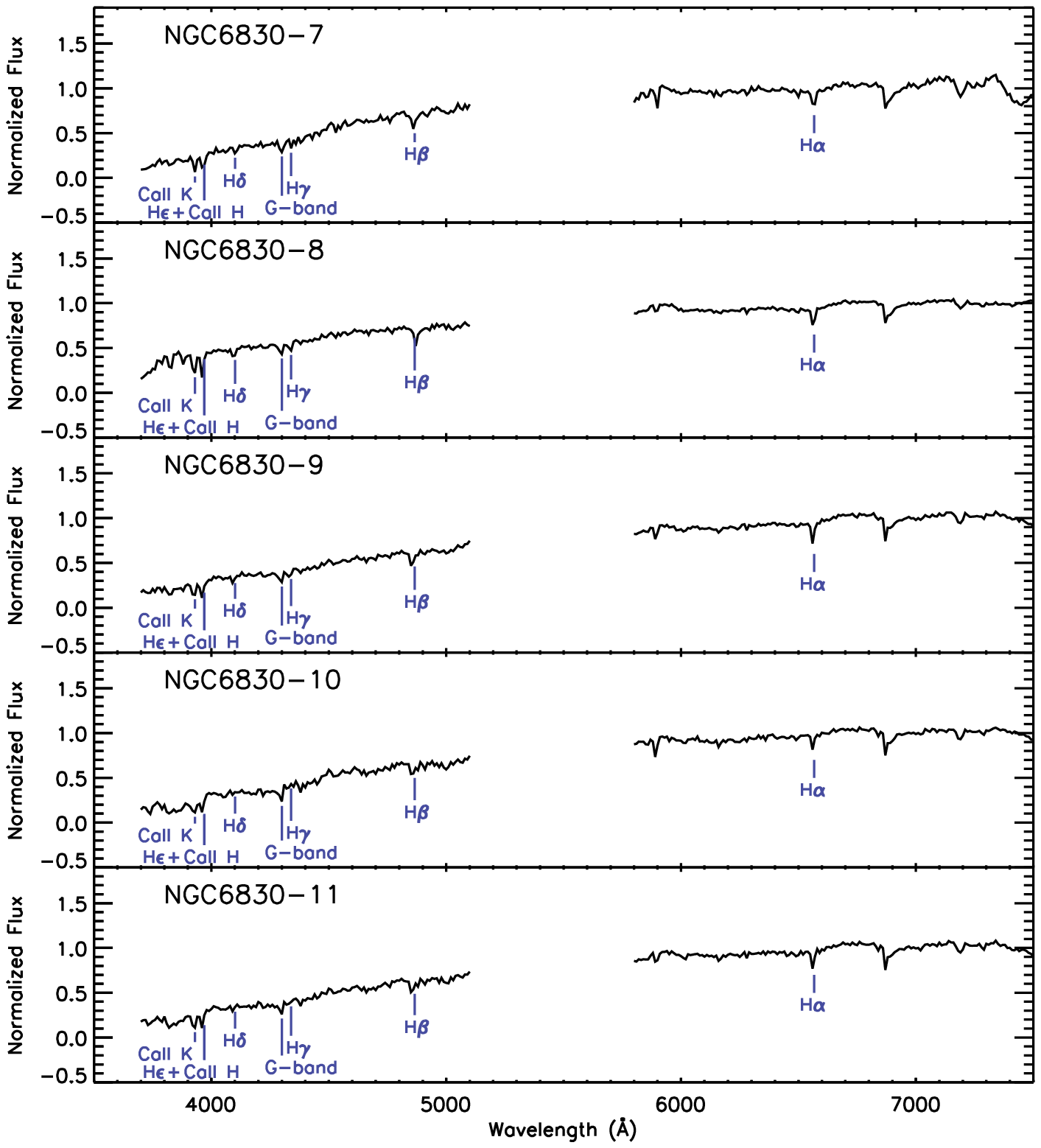}
\caption[]{Lick spectra of Be star candidates}
\label{Lick2}
\end{figure*}

\renewcommand{\thefigure}{\arabic{figure}}

\subsubsection{Confirmed Be Stars} 
(1) NGC\,6830-1: This candidate shows a prominent H$\alpha$ emission line with the EW[H$\alpha$] of $-7$\AA~and
hydrogen absorption lines of H$\beta$, H$\gamma$, H$\epsilon$, H$\delta$. The EW[H$\alpha$] is similar to that of the LOT observation.
It also shows \ion{He}{1}$\lambda$4026, \ion{He}{1}$\lambda$4387, and \ion{He}{1}$\lambda$4471 + \ion{Mg}{2}$\lambda$4481 absorption lines, the features of B-type stars. 
We thus confirmed the candidate as a newly discovered Be star.

(2) NGC\,6830-2: NGC6830\,2 is a known Be star VES\,72, a B6 type star with photoelectric observations \citep{Hoag1965}.
We also detected the H$\alpha$ emission line, with the EW[H$\alpha$] = $-9$\AA.
It shows similar hydrogen and helium/magnesium absorptions lines as in NGC\,6830-1.

(3) NGC\,6830-3: The star shows the H$\alpha$ line in emission with the EW[H$\alpha$] = $-8$\AA. In addition to the same hydrogen and helium/magnesium absorptions lines
detected in NGC\,6830-1 and NGC\,6830-2, it also shows the \ion{He}{1}$\lambda$4009 absorption line. We thus classified the star as a Be star.
The star is 2 mag fainter than NGC\,6830-1 and NGC\,6830-2, implying a late-type B9 stars.

\subsubsection{Non-Be Star Candidates}

(4) NGC\,6830-4: We do not detect the H$\alpha$ emission line for the star, and it shows the \ion{Ca}{2}~K absorption line, indicating the star might be a late-A or F-type star.

(5) NGC\,6830-5: It shows not only the \ion{Ca}{2}~K absorption line, but also the G-band absorption at 4300\AA. The star is classified as a G-type star.

(6) NGC\,6830-6 -- NGC\,6830-11: Compared to NGC\,6830-5, these stars show a stronger G-band absorption, and weaker H$\gamma$ and H$\epsilon$ absorption lines.
We classify NGC\,6830-6 -- NGC\,6830-11 as G-type stars.

\subsection{Infrared Color}
The Be star NGC\,6830-3 shows infrared excess $H-K_{s}$ = 0.316 (Figure~1d), as in many known Be stars.
Such infrared excess can be explained by free-free emission from circumstellar disks \citep{lee11}.
For comparison, we included the fluxes of 3.4 $\micron$~(W1), 4.6 $\micron$~(W2), 12 $\micron$~(W3) and 22 $\micron$~ (W4) from Wide-field Infrared Survey Explorer \citep[WISE;][]{wright2010}, 
and plotted in Figure~5 the spectral energy distribution between 0.4~$\micron$ and 25~$\micron$ of three Be stars (Table~3).
All of the objects have no significant 22~$\micron$ detection, suggesting no presence of warm dust.
Although the detection is only an upper-limit, it seems that NGC\,6830-3 might have a different W3$-$W4 color from NGC\,6830-1 and NGC\,6830-2.
Observations with high resolution and sensitivity are in demand to confirm the 22~$\micron$ detection in NGC\,6830-3 in the future.

\subsection{Optical Variability}
Be stars are also known as photometric and spectroscopic variables.  
Their variabilities show a wide range of timescales from much less than a day \citep{Percy2002} to more than decades \citep{Okazaki1997}.  
Compare to normal B type stars, photometric irregular variation is frequently seen in Be stars \citep{Kourniotis2014}.
There are three types of variability found in Be stars: (1) short-term variability with time scales of days \citep{Hubert1997}; (2) mid-term variability with time scales of
days and weeks with the amplitude of 0.3 mag \citep{Okazaki1997,Hubert1998}; (3) long-term variability with time scales of years and larger amplitudes of $>$ 0.3 mag \citep{Hubert1998}.
Since the PTF R-band light curve data saturated around 14th magnitudes, we adopted the V-band light curve data from the All Sky Automated Survey \citep[ASAS;][]{Pojmanski2002} to investigate the variability of our detected Be stars.
The ASAS is a project to monitor stars brighter than 14 magnitudes to investigate their photometric variability; thus it is suitable to search for long-term variability with large amplitude $>$ 0.3 mag for Be stars.
We chose two stars, star1 and star2 in the same field, as the photometric references to build up the light curves using differential photometry technique.
We only used the data with grade A from ASAS catalog, i.e., best data with photometric uncertainties $<$~0.05 mag in the catalog.
The value $\sigma_{sc}$ was defined as the standard deviation of magnitude difference between the reference and the Be stars, and used to determine the significance of the variability.
The $\sigma_{sc}$ against star1 and star2 of NGC6830-1, NGC6830-2, and NGC6830-3 are 0.15 and 0.14, 0.14 and 0.16, and 0.26 and 0.23 mag,
respectively (Figure~6). 
These $\sigma_{sc}$ values are all comparable with the systematic errors of corresponding magnitudes in ASAS \citep{Pojmanski2002}. Therefore, we
do not detect any significant long-term variability in optical band for these
three Be stars.

\begin{figure}
\epsscale{1}
\plotone{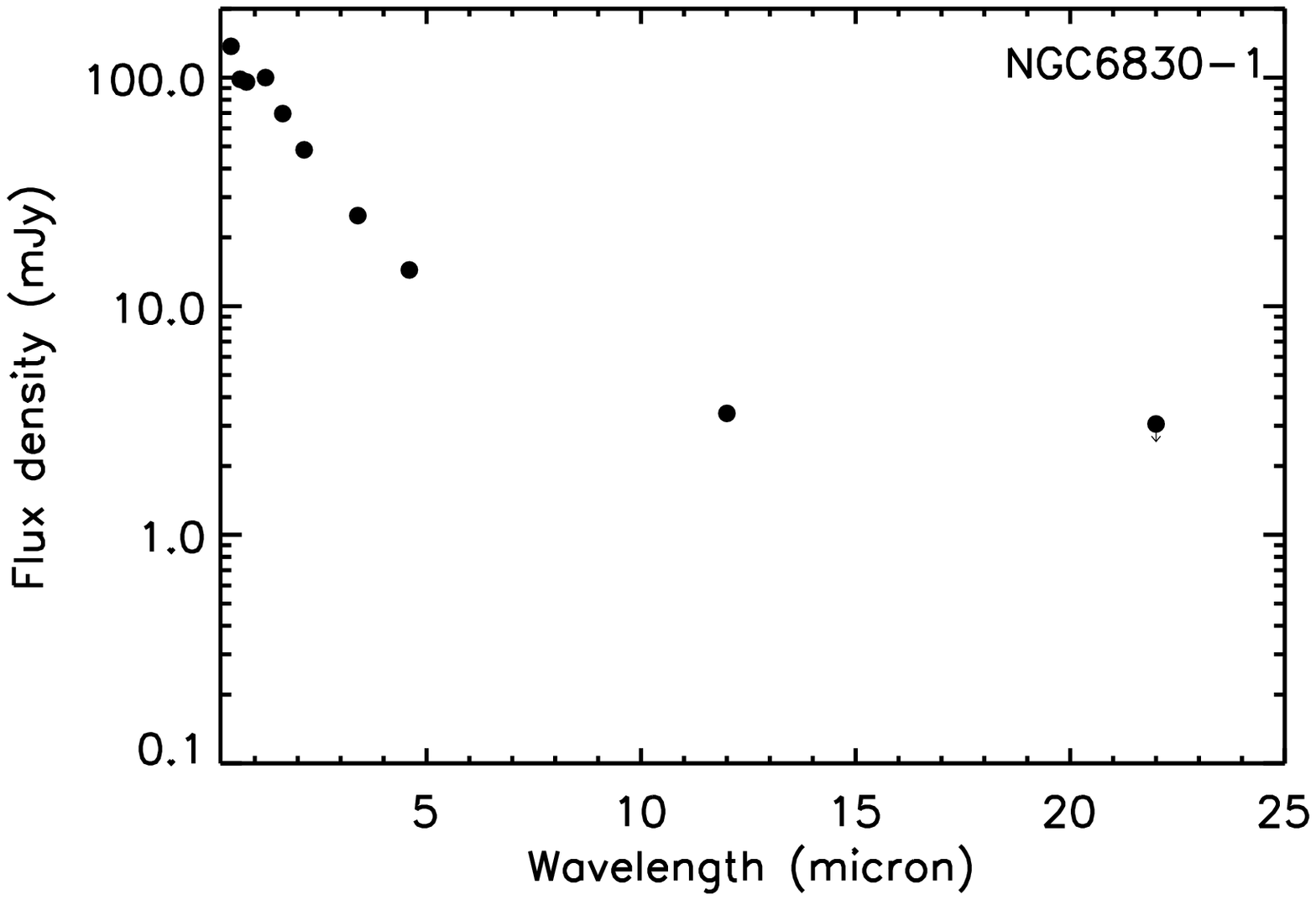}
\plotone{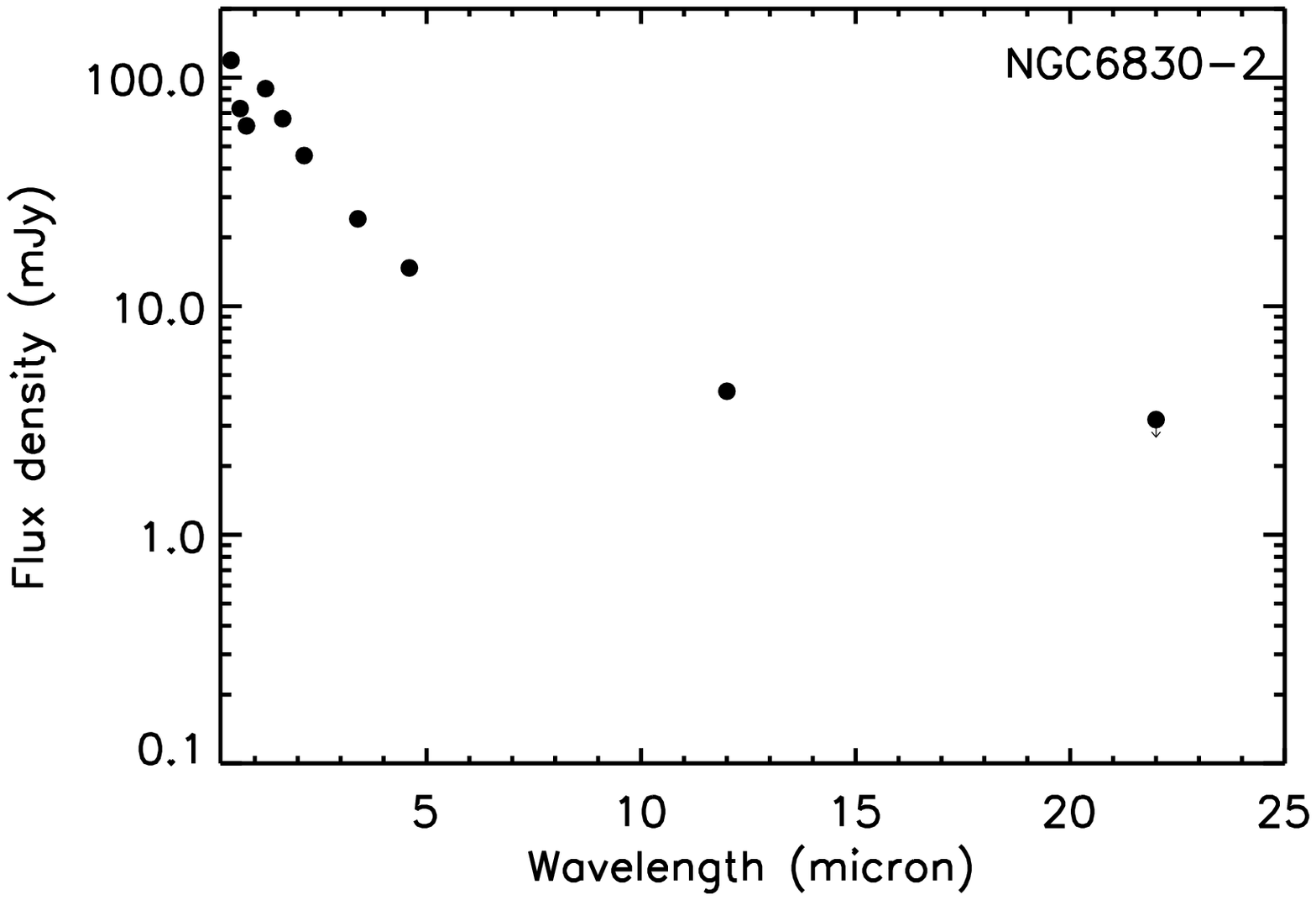}
\plotone{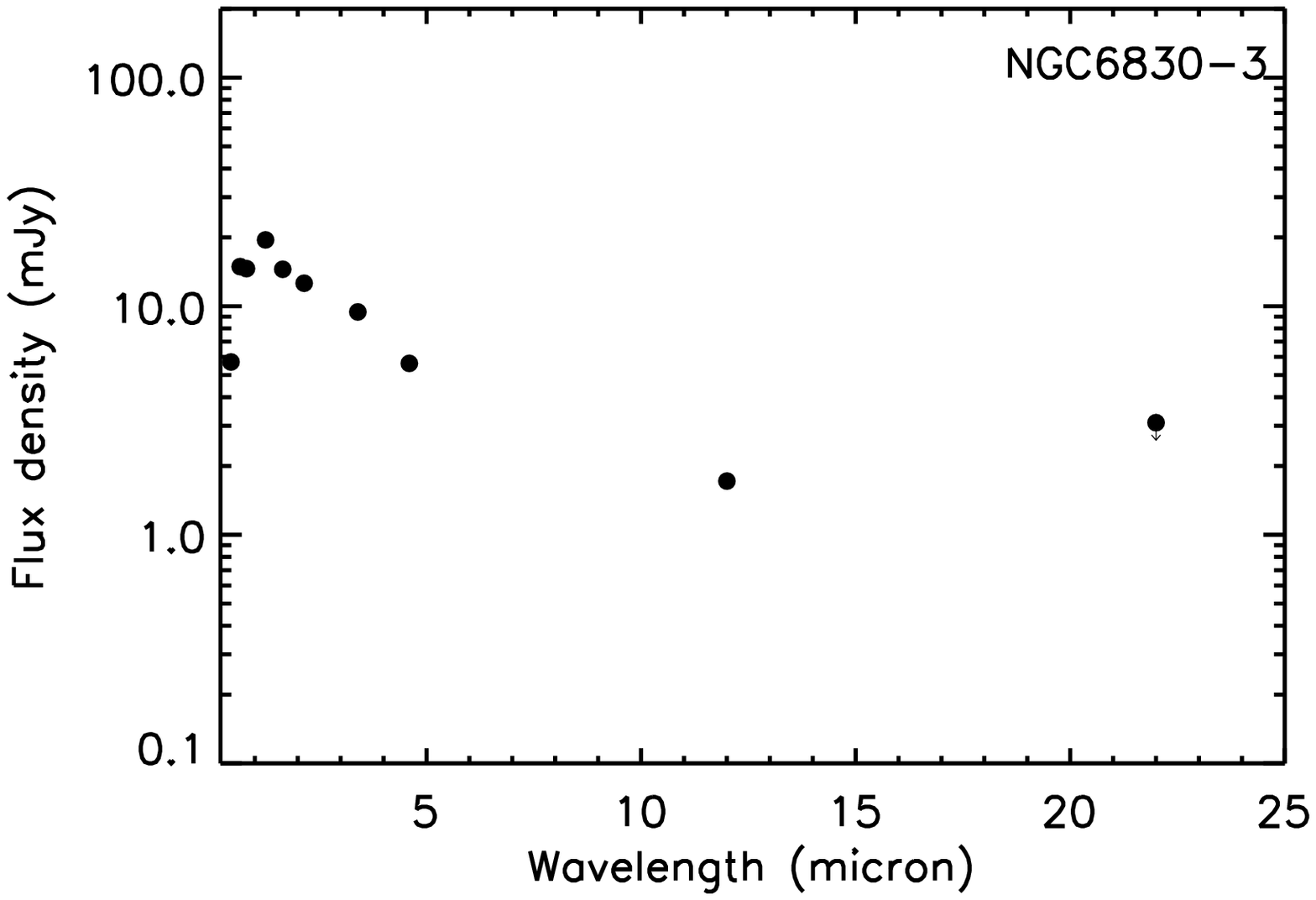}
\caption{Spectral energy distribution between 0.4$\micron$ and 25$\micron$. Detections at 22 $\micron$ for these Be stars are upper-limit. Flux density of 445~nm, 658~nm, and 806~nm are adopted from PPMXL; flux density of 1.25, 1.65, and 2.15 $\micron$
are adopted from 2MASS; 3.4, 4.6, 12, and 22 \micron~flux density are adopted from WISE.}
\label{SED-color}
\end{figure}

\begin{figure}
\epsscale{1}
\plotone{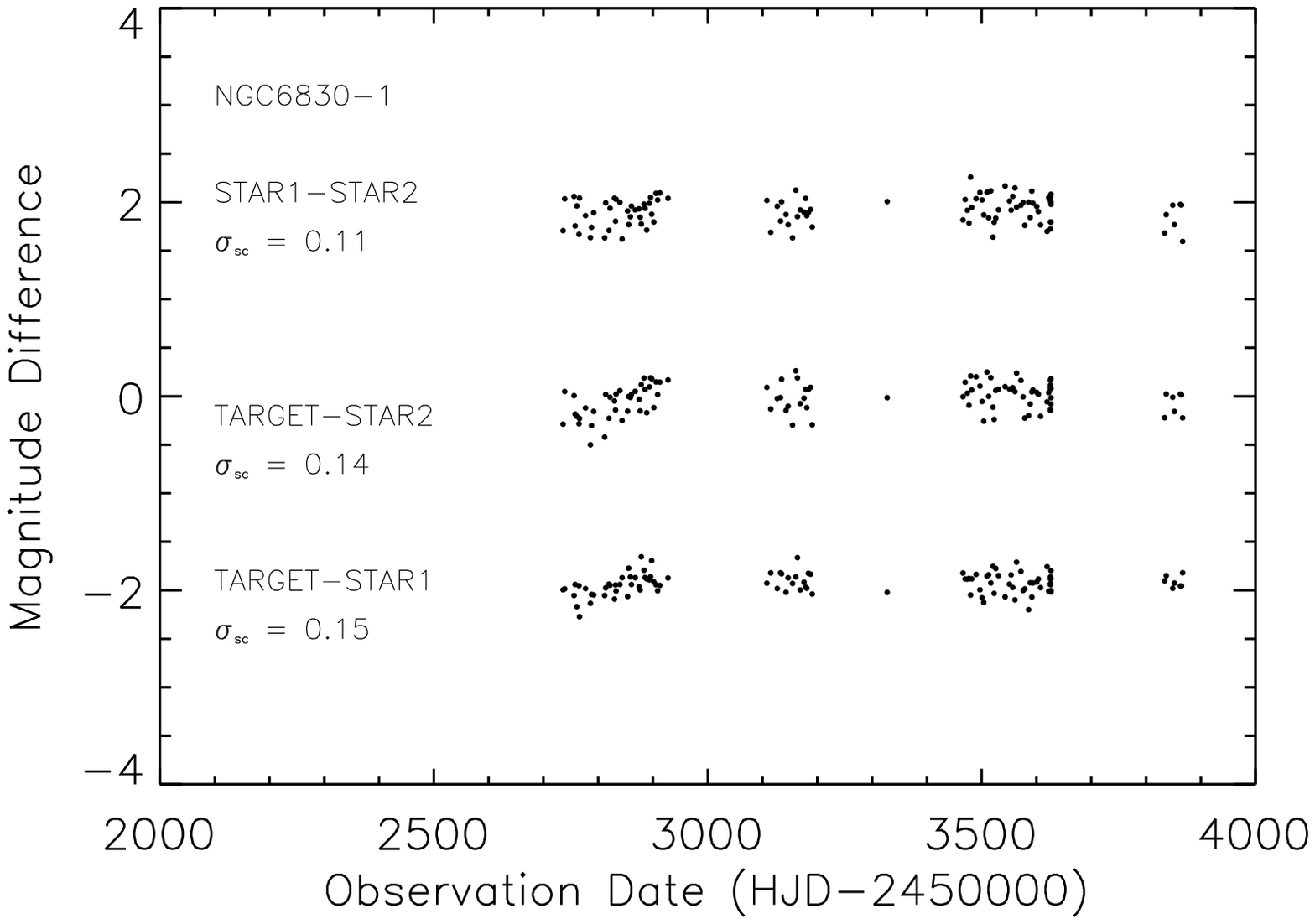}
\plotone{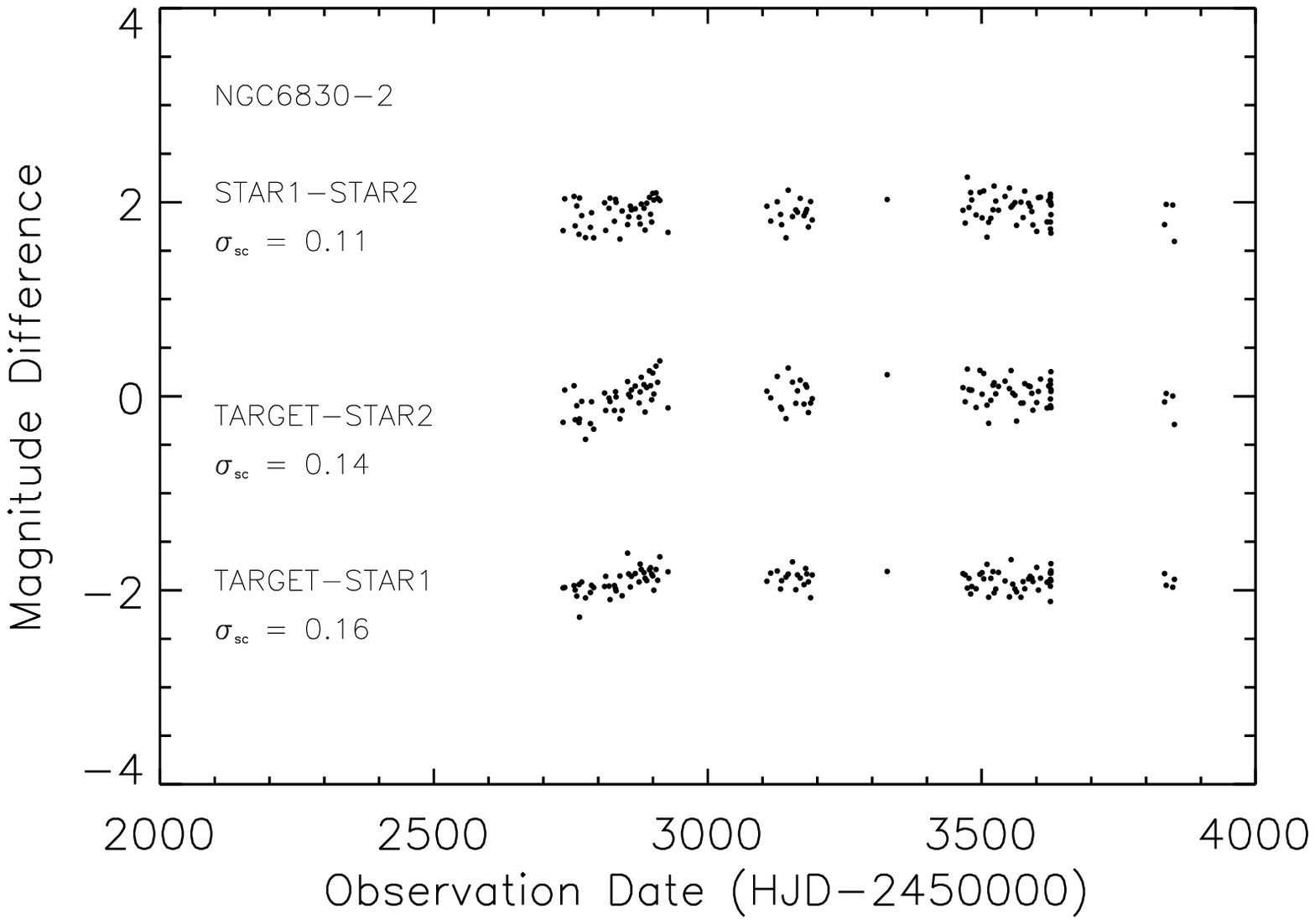}
\plotone{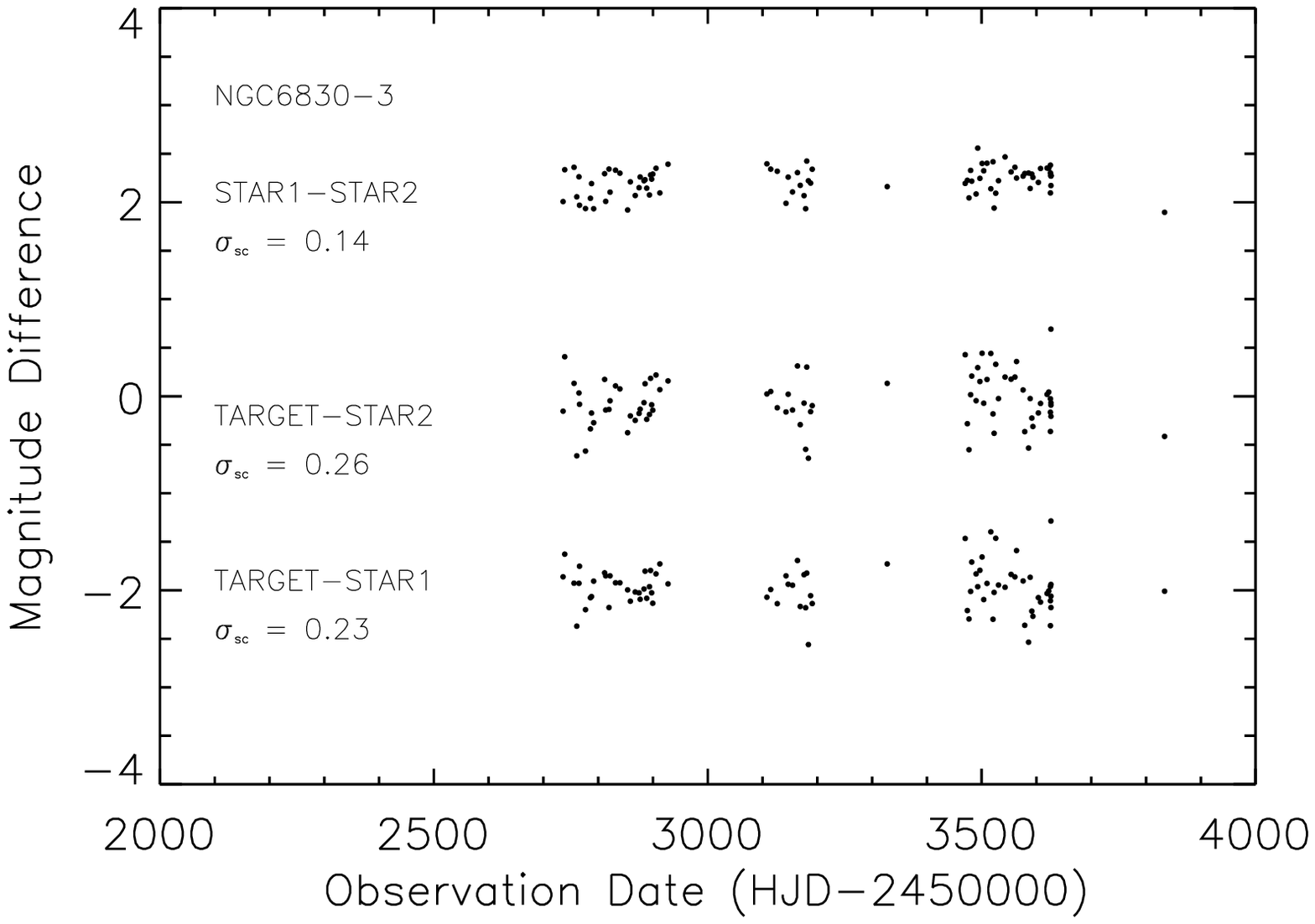}
\caption{Differential photometry of Be stars in NGC\,6830.}
\label{variability}
\end{figure}

\begin{figure}[!ht]
\epsscale{1}
\plotone{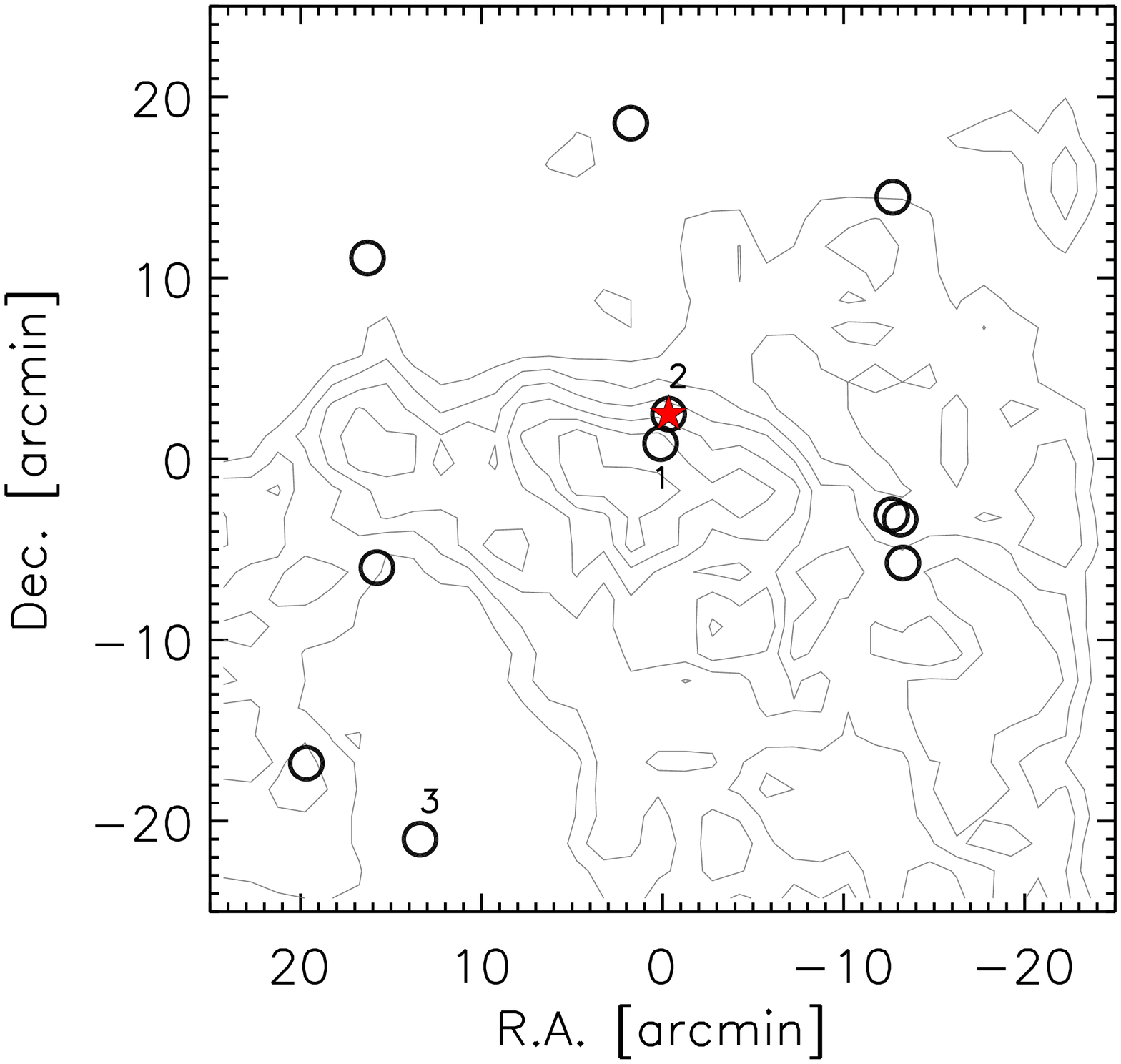}
\caption{Spatial distribution of Be star candidates. The open circles: Be star candidates; red pentagram: the known Be star VES\,72. Contours represent distribution of stars within a field of 50$\arcmin$ $\times$ 50$\arcmin$.}
\label{spatial}
\end{figure}

\begin{figure}[!ht]
\epsscale{1.1}
\plotone{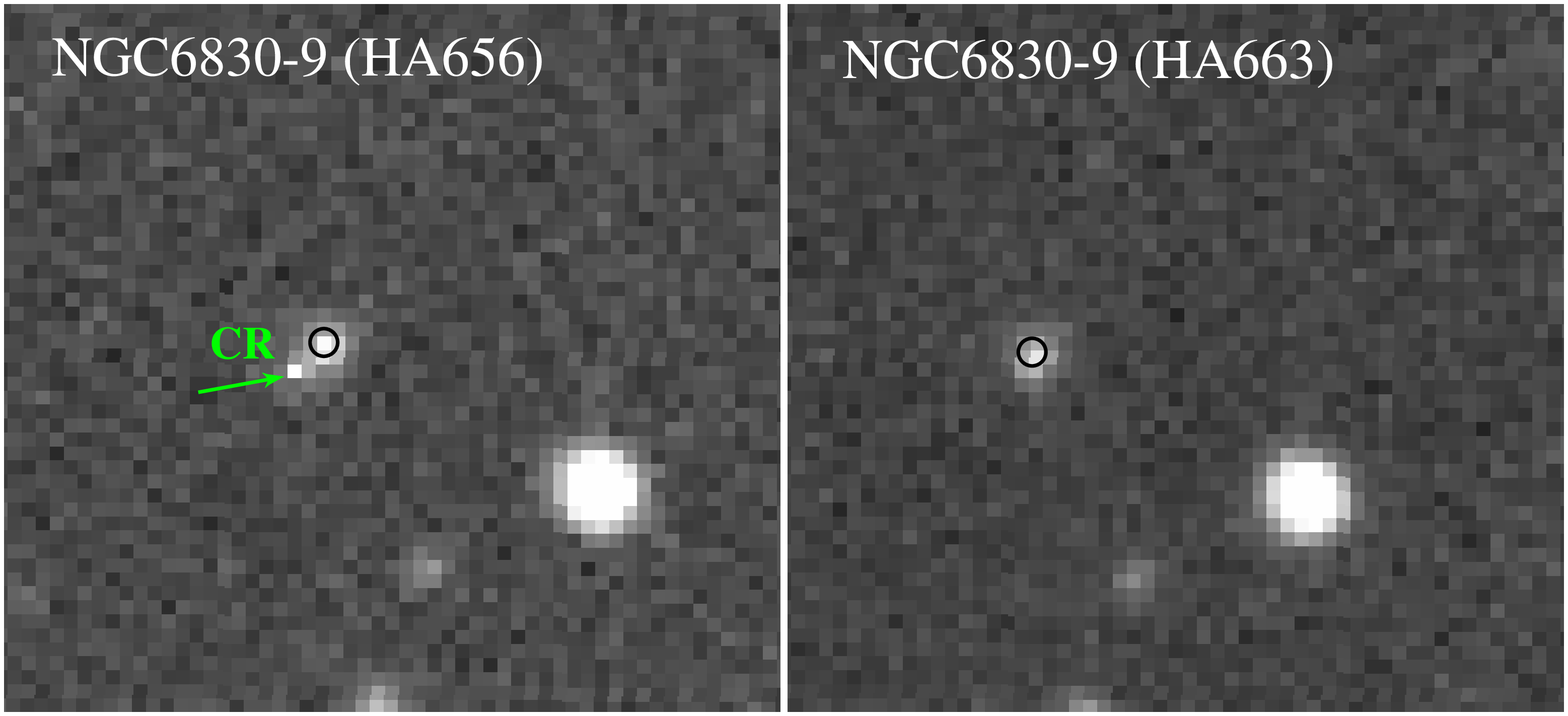}
\caption{HA656 (Left) and HA663 (Right) images of NGC6830-9. Green arrow indicates the residual cosmic-ray. Black open circles represent the position of NGC\,6830-9.}
\label{spatial}
\end{figure}

\begin{deluxetable*}{lrrrrrrrrrr}
\tabletypesize{\small}
\tablecolumns{5} \tablewidth{0pt}\tablecaption{Flux density of Be Stars in NGC\,6830}
\tablehead{
\colhead{ID} &
\colhead{0.445$\micron$} &
\colhead{0.658$\micron$} &
\colhead{0.806$\micron$} &
\colhead{1.25$\micron$} &
\colhead{1.65$\micron$} &
\colhead{2.2$\micron$} &
\colhead{3.4$\micron$} &
\colhead{4.6$\micron$}&
\colhead{12$\micron$}&
\colhead{22$\micron$} \\
\colhead{} &
\colhead{(mJy)} &
\colhead{(mJy)} &
\colhead{(mJy)} &
\colhead{(mJy)} &
\colhead{(mJy)} &
\colhead{(mJy)} &
\colhead{(mJy)} &
\colhead{(mJy)} &
\colhead{(mJy)} &
\colhead{(mJy)}
}
\startdata
NGC\,6830-1  &137 &98.3 &95.7 &100 &69.6 &48.3 &24.9 &14.4 &3.397 &3.053\\
NGC\,6830-2  &119 &73.2 &61.5 &89.5 &66.1 &45.6 &24.1 &14.7 &4.245 &1.717\\
NGC\,6830-3  &5.7 &14.9 &14.6 &19.5 &14.5 &12.6 &9.434 &5.621 &1.717 &3.093
\enddata
\tablecomments{Column 1: ID of targets. Column 2 to 11: Flux density in mJy adopted from PPMXL, 2MASS, and WISE catalog. The detections at 22$\micron$ for three Be stars are upper-limit.}
\end{deluxetable*}

\section{Summary and Discussion}
In summary, we apply PTF's H$\alpha$ imaging photometry to identify 11 Be star candidates in NGC\,6830. 
Three stars have been confirmed as Be stars with intermediate and late type spectra by using the SED-Machine on Palomar 1.5-m telescope, the Hiyoyu spectrograph on Lulin 1-m telescope, and the Kast dual spectrograph on Lick 3-m telescope.
We suggest that there are three Be stars in the cluster NGC\,6830.
The spatial distribution of the Be stars might be caused by gravitational disruption over time. 
We also present the results of the SED-Machine; this study demonstrated that the high efficiency of the SED-Machine can provide rapid observations for Be stars in a comprehensive survey in the future.

We suggest that there are three Be stars in the cluster NGC\,6830.
\citet{McSwain05} studied the fraction of Be stars in eight clusters with an age older than 100~Myr; only four clusters have up to two Be stars.  
These results indicate that old open clusters lack Be stars.
Although our results are consistent with the previous conclusions, it should be noted that either the relaxation process or Galactic external perturbation \citep{Chen2004} might have happened in open clusters.  
As shown in Figure~7, the Be star NGC6830-3 is located in the outer region of the cluster, while NGC6830-1 and NGC6830-2 are located in the central region.
With an age of 125~Mys, the open cluster NGC\,6830 might have a very irregular shape and become loose.
The spatial distribution of Be stars could be the consequence of the relaxation process.
Since \citet{McSwain05} searched Be stars within a fixed angular size, more Be stars could be discovered if a larger size would have been used.

\citet{Dachs1988} proposed an empirical relation between the disk fraction $f_{D}$ and the EW[H$\alpha$] for Be stars:
\begin{equation}
f_{D} \simeq 0.1 \times \frac{EW[H\alpha]}{-30\AA},
\label{discfra}
\end{equation}
where $f_{D}$ is the ratio of disk radiation to total radiation of the star-disk system. 
The EW[H$\alpha$] of the Be stars in NGC\,6830 are all $>$ $-$10\AA, suggesting the disc fraction $f_{D}$ $<$ 0.03.
The result is consistent with \citet{Dachs1988}, who showed that most late-type Be stars (B4 -- B7) have $f_{D}$ $<$ 0.1, while
most early-type Be stars (B1 -- B3) have $f_{D}$ $>$ 0.1.

Several mechanisms might cause long-term variability in Be stars: (1) long-lived strong outbursts \citep{Hubert1998}; (2) decline of brightness due to cooling envelope \citep{Koubsky1997}; (3) weak brightness change related to H$\alpha$ variability. 
The brightness of NGC\,6830-1 and NGC\,6830-2 allow us to investigate the variability with the amplitude larger than 0.15 mag using ASAS data set.
We do not discover any significant variability for NGC\,6830-1 and NGC\,6830-2.
This is consistent with the previous studies; \citet{Hubert1998} showed that only 12\% of B6-type Be stars have the variability with the amplitude of 0.12 to 0.3 mag.
Because large amplitude caused by strong outbursts usually seen in earliest Be stars, it is not surprised that we do not detect variability in these Be stars.

Finally, as a pilot project to search for Be stars in clusters with different ages, we reviewed our selection criteria for NGC\,663 \citep{Yu2015} and this work.
Eight of 11 Be candidates are classified as A- or G-type stars; most of them have marginal H$\alpha$ detection in the HA663$-$HA656 color diagram (Figure~2). 
This misidentification can be improved by using the 2MASS color-color diagram. 
As shown in Figure~1d, if we make the selection region (dashed line) to have similar range as the known Be stars,
most G-type stars would have been ruled out. Thus, we will change our selection criteria in our future project.
Moreover, it should be noted that the star NGC6830-9 has large H$\alpha$ excess in Figure~2, while the optical spectrum suggested a G-type star with the H$\alpha$ absorption line.
This is caused by the contamination of residual cosmic-ray hitting on the target region, which is not easy to be eliminated.
Procedures of removing cosmic-ray around science targets will be performed carefully in the future project.

\acknowledgments
We thank to the referee for her/his constructive comments.
We are also grateful to the staff of Lulin and Lick Observatory for helping the observations. 
This work is supported in part by the National Science Council, and Ministry of Science and Technology of Taiwan under grants MOST 104-2119-M-008-024 (W.-H.I.), MOST 103-2112-M-008-024-MY3 (W.-P.C.), MOST 104-2112-M-008-012-MY3 (C.-C.N.), and NSC 103-2917-I-564-004 (P.-C.Y.). 
The Lulin Observatory is funded by the Ministry of Science and Technology of Taiwan,
and operated by National Central University of Taiwan. The Lick Observatory is funded by Google Inc., and operated by the University of California.
This publication makes use of data products from the Two Micron All Sky Survey, which is a joint project of the University of Massachusetts and the Infrared Processing and Analysis Center/California Institute of Technology, funded by the National Aeronautics and Space Administration and the National Science Foundation. This publication makes use of data products from the Wide-field Infrared Survey Explorer, which is a joint project of the University of California, Los Angeles, and the Jet Propulsion Laboratory/California Institute of Technology, funded by the National Aeronautics and Space Administration.

\end{CJK*}
\end{document}